# Remarkable antibacterial activity of reduced graphene oxide functionalized by copper ions

*Yusong Tu[†,\*], Pei Li[†], Jiajia Sun, Jie Jiang, Fangfang Dai, Yuanyan Wu, Liang Chen, Guosheng Shi, Yanwen Tan[\*], Haiping Fang*

**Despite long-term efforts for exploring antibacterial agents or drugs, it remains challenging how to potentiate antibacterial activity and meanwhile minimize toxicity hazards to the environment. Here, we experimentally show that the functionality of reduced graphene oxide (rGO) through copper ions displays selective antibacterial activity significantly stronger than that of rGO itself and no toxicity to mammalian cells. Remarkably, this antibacterial activity is two orders of magnitude greater than the activity of its surrounding copper ions. We demonstrate that the rGO is functionalized through the cation–π interaction to massively adsorb copper ions to form a rGO–copper composite in solution and result in an extremely low concentration level of surrounding copper ions (less than ~0.5 μM). These copper ions on rGO are positively charged and strongly interact with negatively charged bacterial cells to selectively achieve antibacterial activity, while rGO exhibits the functionality to not only actuate rapid delivery of copper ions and massive assembly onto bacterial cells but also result in the valence shift in the copper ions from $Cu^{2+}$ into $Cu^{+}$ which greatly enhances the antibacterial activity. Notably, this functionality of rGO through cation–π interaction with copper ions can similarly achieve algaecidal activity but does not exert cytotoxicity against neutrally charged mammalian cells. The remarkable selective antibacterial activity from the rGO functionality as well as the inherent broad-spectrum-antibacterial physical mechanism represents a significant step toward the development of a novel antibacterial material and reagent without environmental hazards for practical application.**

*Prof. Y. Tu, J. Sun, Dr. Y. Wu*
*College of Physics Science and Technology, Yangzhou University*
*Jiangsu, 225009, China*
*E-mail: ystu@yzu.edu.cn; jjsun_yzu@163.com; yywu@yzu.edu.cn*
*Prof. Y. Tu*
*Key Laboratory of Polar Materials and Devices Ministry of Education, School of Physics and Electrical Science*
*East China Normal University, Shanghai 200241, China*
*E-mail: ystu@clpm.ecnu.edu.cn*
*P. Li, Prof. Y. Tan*
*State Key Laboratory of Surface Physics, Multiscale Research Institute of Complex Systems*
*Department of Physics, Fudan University Shanghai 200433, China*
*E-mail: peili92@alu.fudan.edu.cn; ywtan@fudan.edu.cn*
*J. Jiang, Prof. H. Fang*
*Division of Interfacial Water and Key Laboratory of Interfacial Physics and Technology, Shanghai Institute of Applied Physics, Chinese Academy of Sciences, Shanghai 201800, China*
*E-mail: jiangjie@sinap.ac.cn; fanghaiping@sinap.ac.cn*
*F. Dai, Prof. L. Chen*
*Department of Optical Engineering, Zhejiang A&F University, Lin'an 311300, China*
*E-mail: fangfangdai94@126.com; liang_chen05@126.com*
*Prof. G. Shi*
*Shanghai Applied Radiation Institute, Shanghai University, Shanghai 200444, China*
*E-mail: gsshi@shu.edu.cn*
*Prof. H. Fang*
*School of Science, East China University of Science and Technology, Shanghai 200237, China*
[†]These authors contributed equally to this work.

# 1. Introduction

How to selectively potentiate antibacterial activity and meanwhile minimize toxicity hazards to the environment is a major and vital question for improving human and animal health and agricultural yields [1]. Antibacterial drugs or antibiotics were introduced more than 70 years ago, however, their widespread uses in healthcare and agriculture have resulted in the accumulation of antibiotics in the environment [2]. This has become an increasingly common health threat due to the emergence and spread of bacterial resistance in the environment [3]. Also, copper antibacterial activity was recognized in ancient times, and its ions and complexes have been shown to possess broad-spectrum antimicrobial [4], antifungal and antiviral activity [5]. However, in excess, copper ions become toxic and adversely pose hazards to the environment, especially to the aquatic environment [6]. Despite long-term efforts on counteracting the increasing resistance of bacteria by decreasing the inappropriate use of antibiotics as well as researching and developing new antibacterial materials or agents [7], there is an urgent need yet a challenge for achieving the selective, highly-efficient, and environmentally-friendly antibacterial activity.

Graphene is a 2D single-atom-thick nanomaterial with hexagonally-assigned $sp^2$-bonded carbon [8], and its unique structure can result in excellent physicochemical properties for various applications [9]. Previous researches have shown that graphene can present antibacterial activity [10], and the activity mainly originate from its characteristic interactions with bacterial cells, i.e., cutting membranes or extracting large amounts of phospholipids from membranes or covering onto membranes [10a, 10b], in comparison with the translocating or penetrating membranes of other 0D or 1D nanomaterials [11]. Many efforts have also been made to study the antibacterial activity of functionalized graphene, but the enhancement of this activity was found to be less than twofold in comparison with graphene [12]. For example, graphene functionalized by conjugating polymers [12a] and with controlled orientational alignment [12b] had antibacterial activity enhanced by only 1.30- and 1.46-fold, respectively.

Here, we experimentally report an unexpectedly amplified selective antibacterial activity of reduced graphene oxide (rGO) functionalized by only copper ions, without toxicity to mammalian cells. The antibacterial activity is significantly stronger than that of rGO itself and two orders of magnitude greater than that of the surrounding copper ions. We have performed density functional theory (DFT) calculations and molecular dynamics (MD) simulations showing that rGO is functionalized through cation–π interaction to massively adsorb copper ions to form a rGO–copper composite and result in an extremely low concentration level of surrounding copper ions (less than ~0.5 μM). These copper ions on rGO are positively charged and strongly interact with negatively charged bacterial cells to selectively achieve antibacterial activity. Importantly, along with our field-emission scanning electron microscopy (FE-SEM) and X-ray photoelectron spectrometer spectra (XPS) experiments, these results reveal the inherent broad-spectrum-antibacterial physical mechanism on the functionality of rGO through cation-π interaction with copper ions, i.e., the rGO functionality not only actuates rapid delivery of copper ions and massive assembly onto bacterial cells but also results in the valence shift in the copper ions from $Cu^{2+}$ into $Cu^+$, which greatly enhances the antibacterial activity. Notably, this functionality of rGO through copper ions can similarly achieve algaecidal activity but does not exert cytotoxicity against neutrally charged mammalian cells. Our results indicate that this rGO functionality has taken the most advantage of the charge difference between the cells to achieve the remarkable selective antibacterial activity and meanwhile avoiding environmental hazards.

# 2. Results and discussions

**2.1 Antibacterial Activities from rGO Functionality through Copper Ions.**

We prepared rGO suspensions from natural graphite powder by using a modified Hummers method and electron-beam irradiation reduction (see the Experimental Section for details). The as-prepared rGO was mainly a single-layer sheet with average lateral dimension of ~10 μm (Supporting Information, Figure S2). We added a small amount of the rGO suspension to copper sulphate (CuSO$_4$) solution to form a rGO–copper composite solution. We used a conventional colony-forming cell assay to analyze antibacterial activity against Gram-negative *Escherichia coli (E. coli)* and Gram-positive *Staphylococcus aureus (S. aureus)*. As illustrated in **Figure 1A**, after incubation with the rGO–copper composite solution (40 μg mL$^{-1}$ rGO and 100 μM CuSO$_4$), bacterial cells are present in only three and eight colonies for *E. coli* and *S. aureus* respectively, demonstrating remarkable antibacterial activity. By contrast, after incubation with rGO itself (40 μg mL$^{-1}$), bacterial cells are present in dense colonies similar to those in the control experiment, indicating that the rGO hardly provides antibacterial activity. This is consistent with previous experimental results obtained for GO with similar lateral dimensions of 5–20 μm and at concentrations lower than 10$^3$ μg mL$^{-1}$ [13]. Thus, the rGO–copper composite solution has significantly stronger antibacterial activity than the rGO solution itself. To confirm that this enhanced antibacterial activity results essentially from the rGO–copper composite, we investigated the antibacterial activity of surrounding copper ions in the composite solution. Accordingly, the rGO–copper composite was filtered and removed, and the copper ions in the residual solution were measured and presumed to be the surrounding copper ions in the composite solution. Notably, the concentration level of surrounding copper ions is extremely low (lower than ~0.5 μM, detected with inductively coupled plasma optical emission spectrometer (ICP-OES); Supporting Information, Figure S1), which is at the same order of magnitude of water quality criteria [14]. This indicates the functionality of rGO to strongly adsorb copper ions to form a rGO–copper composite. Furthermore, because 0.5 μM is too low a concentration for accurate quantification of antibacterial activity, we postulated approximately 3 μM to be the reference concentration of the surrounding copper ions. For this concentration, the colonies of CuSO$_4$-incubated bacterial cells are slightly sparser than those in the control experiment, indicating weak antibacterial activity (Figure 1A). By comparison, the rGO–copper composite exhibits antibacterial activity at least 220 times higher than that of the surrounding copper ions in the solution (Figure 1B upper). Similarly, results demonstrated that this functionality of rGO through copper ions likewise displays significant antibacterial activity against *S. aureus* ant this activity is ~27 times higher than that of the surrounding copper ions in the solution (Figure 1B down). (Supporting Information, Figure S4 for all three parallel antibacterial experiments).

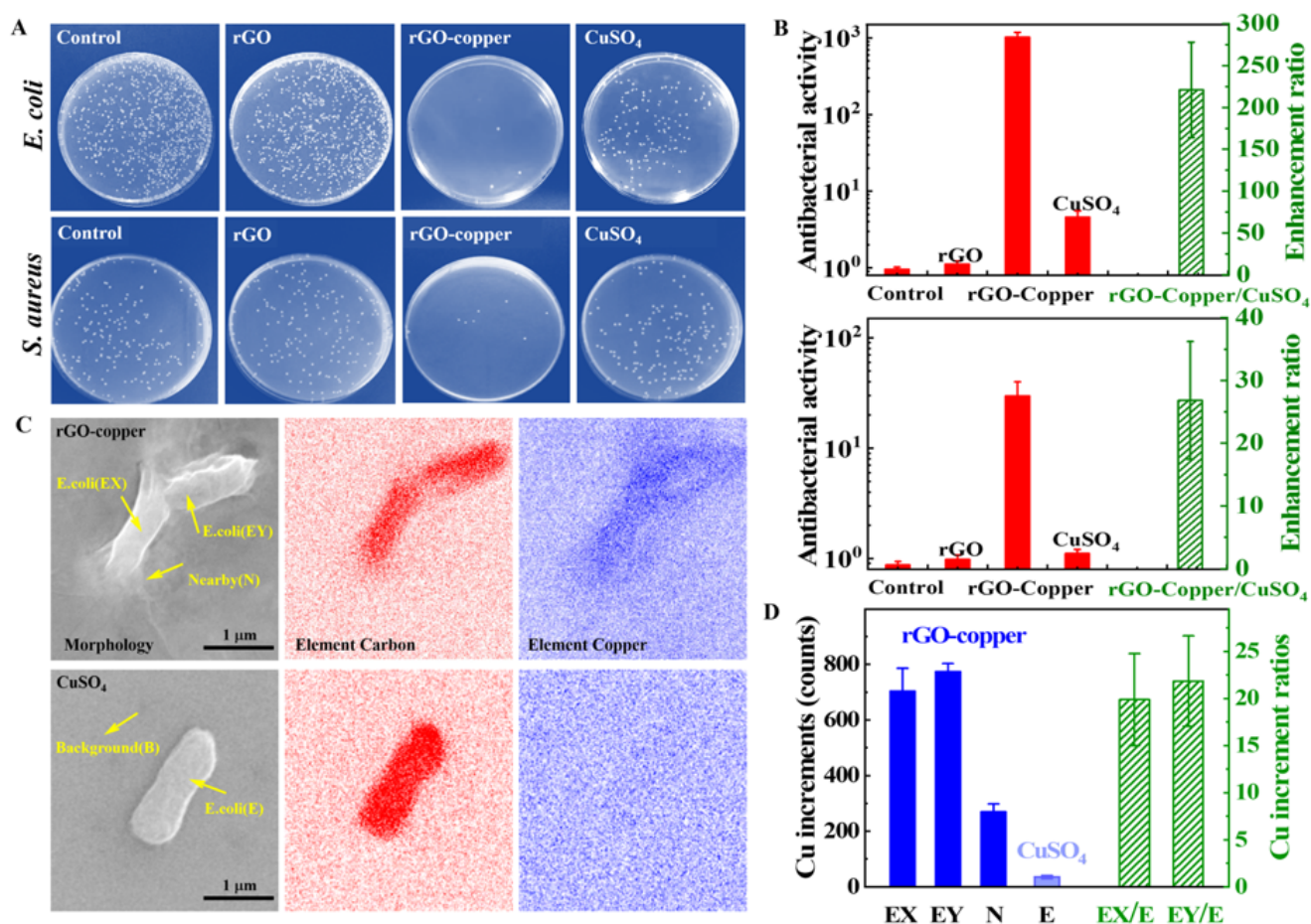

**Figure 1.** Remarkable antibacterial activity of rGO functionalized by copper ions. (A) Images of colony-forming cell assays against *E. coli* (upper row) and *S. aureus* (following row). Bacterial cells were incubated with deionized water as the control, with rGO solution (40 μg mL$^{-1}$), with rGO–copper composite solution (40 μg mL$^{-1}$ rGO and 100 μM copper sulphate [CuSO$_4$]), and with CuSO$_4$ solution (3 μM). The CuSO$_4$ solution (3 μM) was employed as a reference for the surrounding copper ion concentration in the rGO–copper composite solution (determined to be less than ~0.5 μM; Supporting Information, Figure S1). (B) Antibacterial activity (left axis) against *E. coli* (upper) and *S. aureus* (down), represented by the ratio of the colony number in the control experiment to the colony numbers in antibacterial experiments, and enhancement ratio (right axis) of rGO–copper antibacterial activity relative to the antibacterial activity of the surrounding copper ions (3 μM). The error bars indicate standard deviations in experiments with three samples. (C) FE-SEM morphology images and elemental mappings of carbon and copper for the rGO–copper-incubated and CuSO$_4$-incubated *E. coli* cells. The yellow arrows indicate the regions marked by *E. coli* (EX), *E. coli* (EY), nearby (N), *E. coli* (E), and background (B), from which data were collected to quantify the Cu content. (D) Increments of Cu content (left axis) on the rGO–copper-incubated cells (EX and EY) and CuSO$_4$-incubated cells (E) over the background Cu content (B) as well as the ratio (right axis) of the Cu increment on the cells (EX and EY) to the increment on the cell (E). The average Cu content and errors were evaluated using energy-dispersive X-ray spectroscopy (EDS) measurement performed at five random loci around the arrow-indicated regions in A (Supporting Information, Figure S3).

## 2.2 Elemental Cooper Mapping by FE-SEM.

We adopted FE-SEM to test whether this antibacterial activity originates from the functionality of rGO to adsorb copper ions and even deliver them onto bacterial cells. The elemental carbon mappings presented in Figure 1C are consistent with the morphologies of the corresponding bacterial cells, mainly because of the high C content of the cells. In particular, the rGO is depicted to cover and even wrap the rGO–copper-incubated cells in the morphology image, and these profiles are also consistent with the C mappings. In contrast, the elemental copper mapping of the rGO–copper-incubated cells has a consistent profile with the morphologies of cells covered and wrapped by rGO. The signal intensity for Cu is even comparable to the signal intensity for C from the bacterial cells. However, only background noise signals for Cu are present on the mappings of the CuSO$_4$-incubated cells at the corresponding copper ion concentration. Figure 1D illustrates that the increments of the Cu contents on the rGO–copper-incubated cells over the background Cu content are approximately 20-fold larger than the increments on the CuSO$_4$-incubated cells. These results clearly demonstrate the functionality that the rGO massively adsorbs copper ions and further delivers and assembles copper ions onto bacterial cells to greatly enhance the antibacterial activity.

## 2.3 Antibacterial Activities from the Functionality of Graphene Oxide Quantum Dots (GOdot) through Copper Ions.

We examined the functionality of GOdot (around ten nanometers, confirmed by AFM, Supporting Information, Figure S2) for antibacterial activity. A small amount of GOdot suspension were added to CuSO$_4$ solution to form a GOdot–copper composite solution, and the antibacterial experiments were performed under the same concentration conditions as rGO–copper composite (40 μg mL$^{-1}$ GOdot and 100 μM CuSO$_4$). As shown in **Figure 2A** and 2B, despite different lateral sizes, the GOdot presents consistent functionality with rGO, through copper ions to achieve remarkable antibacterial activity. Nevertheless, in contrast to rGO, the GOdot–copper composite exhibits significantly stronger antibacterial activities, and the enhancement ratios of the GOdot–copper antibacterial activity are even up to ~1×10$^6$ and ~5×10$^3$ times, respectively against *E. coli* and *S. aureus*.

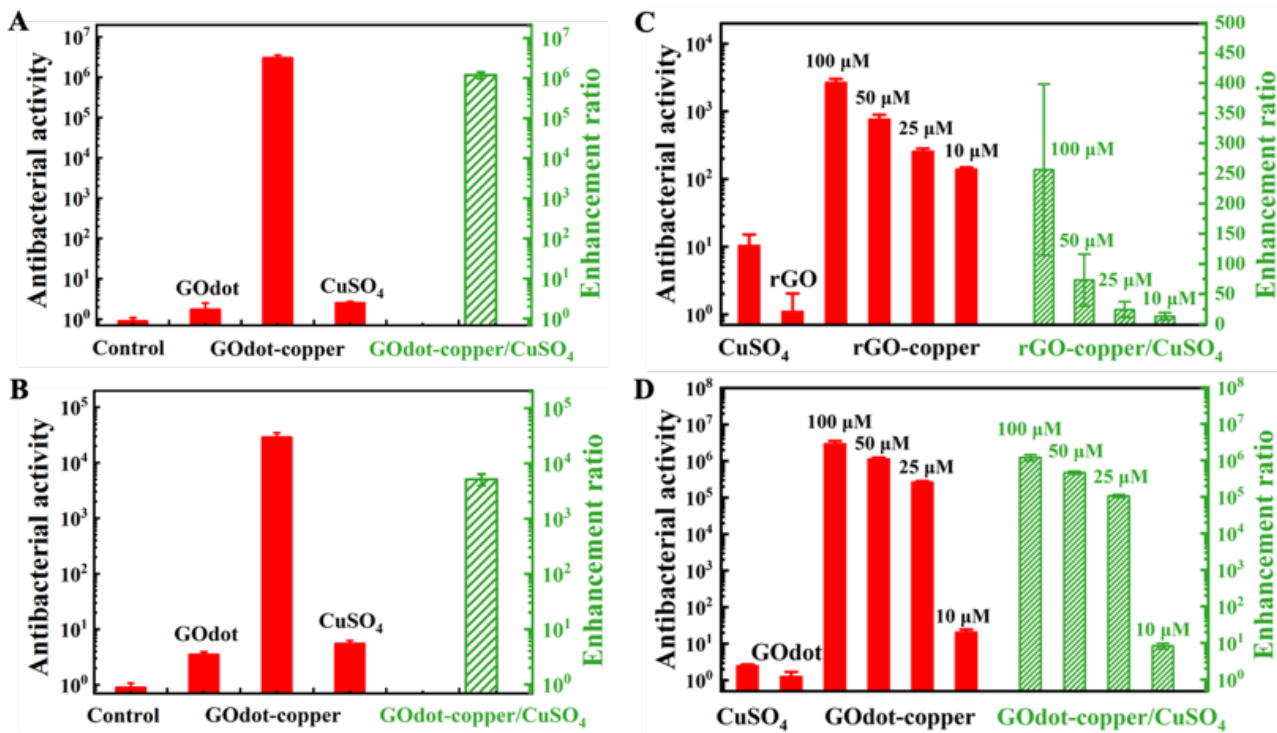

**Figure 2.** (A) and (B) Antibacterial activities (left axis) of GOdot functionalized by copper ions and enhancement ratios (right axis) of GOdot–copper antibacterial activity relative to the antibacterial activity of the surrounding copper ions (3 μM), against *E. coli* (A) and *S. aureus* (B). The concentration conditions are the same as in Figure 1, besides 40 μg mL$^{-1}$ GOdot. The CuSO$_4$ solution (3 μM) was also postulated as a reference for the surrounding copper ion concentration in the GOdot–copper composite solution. (C) and (D) Tendencies of the antibacterial activities and the enhancement ratios against *E. coli* with respect to copper ion concentrations (100 μM, 50 μM, 25 μM, 10 μM). The error bars indicate standard deviations in experiments with three samples.

Unexpectedly, the GOdot functionality through copper ions exhibits at least more than two orders of magnitudes larger enhancement ratios of antibacterial activities than the rGO functionality. We attribute this effect to the drastically increased number concentration of GOdot under the same mass concentration with rGO sheets. From a simple estimation (Supporting Information, PS3), one large rGO sheet (~10 μm) corresponds to $10^6$ nanosheets of Godot (~10 nm) under the same mass condition, i.e., compared with the rGO, the GOdot can have around $10^6$ times higher probability to interact with bacterial cells under the same mass concentration and consequently brings about the unexpected enhancement ratios of the antibacterial activity. This indicates that the lateral sizes have no essential influence on the rGO functionality and small lateral sizes may provide the significant advantage for further boosting up its enhancement ratios of antibacterial activities.

Furthermore, we performed antibacterial experiments under various copper ion concentrations. Figure 2C and 2D shows the antibacterial activity against *E. coli* from rGO and GOdot functionality with respect to the CuSO$_4$ concentrations as well as the relevant enhancement ratios. While the enhancement ratios present a decrease tendency as the ion concentrations are reduced, the rGO functionality through copper ions under 10 μM still presents the enhancement ratio of 13 times, and the GOdot functionality through copper ions under 25μM displays the enhancement ratio of larger than $10^5$ times. This confirms the validity of the functionality to achieve the remarkable antibacterial activity even under the very low concentration of copper ions.

**2.4 Characterization of rGO–copper Composite and Molecular Mechanism Analysis.**
We used XPS to characterize the rGO–copper composite as well as to measure $CuSO_4$, $Cu_2SO_4$, and rGO only as the references. **Figure 3A** shows their Cu $2p_{3/2}$ XPS spectra. The $CuSO_4$ presents a $Cu^{2+}$ characteristic peak at ~935.0 eV, consistent with the previous XPS spectra of $CuSO_4$ [15]; the $Cu_2SO_4$ presents a $Cu^+$ characteristic peak at ~932.6 eV, consistent with previous XPS spectra of $Cu_2O$ or $Cu_2S$ in the NIST XPS Database despite no data of $Cu_2SO_4$ in the Database [15]. The rGO does not include any copper element and thus presents only flat noise signals. Notably, in contrast, all three samples of rGO–copper composite display the broad overlapping peaks, clearly indicating the $Cu^+$ characteristic peaks near ~932.6 eV besides $Cu^{2+}$ characteristic peaks at ~935.0 eV. This indicates that the copper ions in $CuSO_4$ solution present the valence shift from +2 to +1 after being adsorbed on rGO to form rGO–copper composite.

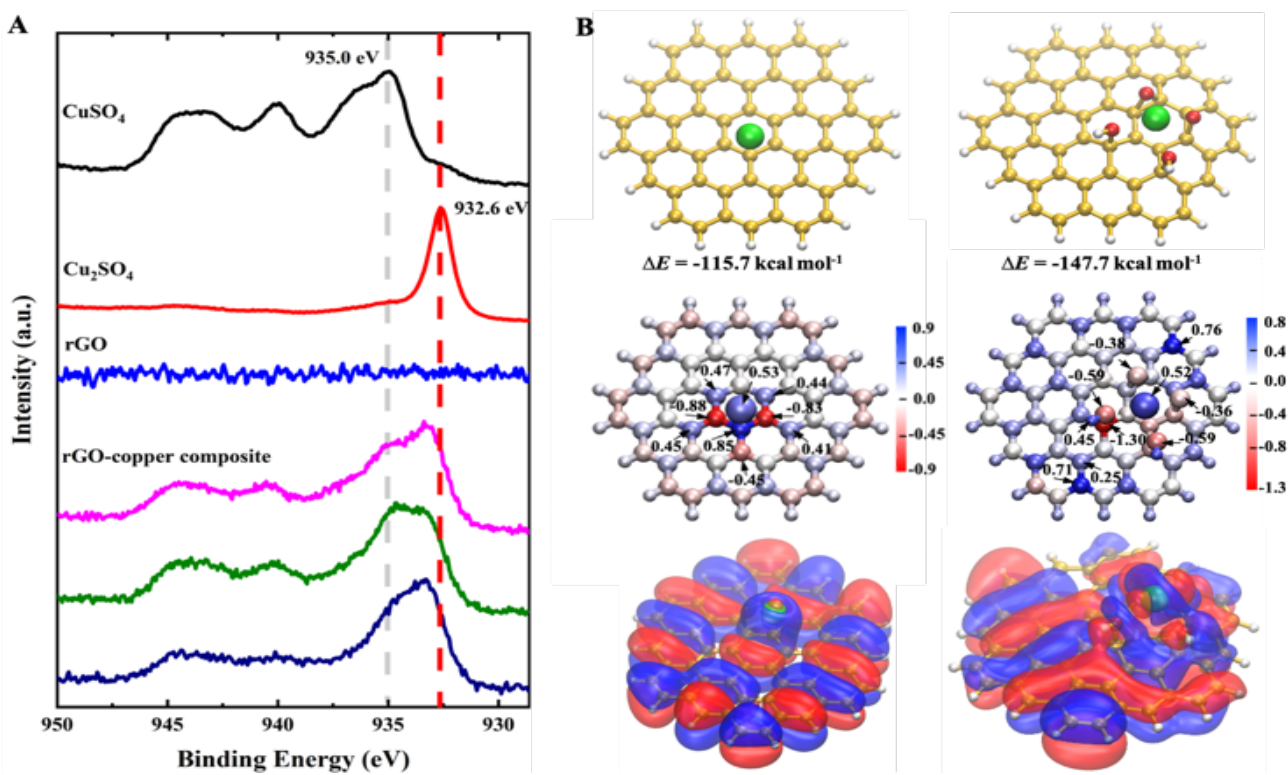

**Figure 3.** Adsorption interaction between rGO and copper ions as well as the rGO–copper composite. (A) Cu $2p_{3/2}$ X-ray photoelectron spectrometer spectra (XPS) of three samples of rGO–copper composite as well as $CuSO_4$, $Cu_2SO_4$, and rGO only. The grey and red dashed lines mark peaks at ~935.0 eV and ~932.6 eV corresponding to the $Cu^{2+}$ and $Cu^+$ characteristics, respectively. (B) The optimized geometries (upper) of copper ion with the valence of +2 on both graphene and GO sheets corresponding to the unoxidized and oxidized regions of rGO, respectively, as well as the relevant Mulliken charge distributions (middle) and highest occupied molecular orbitals (HOMO) (down). $\Delta E = E_{rGO–copper} - E_{copper} - E_{rGO}$, where $\Delta E$ is the binding energy of copper ion adsorption onto rGO, $E_{rGO–copper}$, $E_{copper}$ and $E_{rGO}$ are the total energies of optimized rGO–copper composite, copper ion and rGO, respectively. The rGO is shown as bonded spheres (C: yellow, O: red, and H: white) and copper ion, as a green sphere. The red and blue represent the negative and positive charged atoms. The electron density was plotted for iso values of 0.005 a.u. with the red and blue colors denoting the opposite signs.

DFT calculations were performed to understand the adsorption interaction between rGO and copper ions in rGO–copper composite at the B3LYP/6-31G(d) level of theory. [16] We have optimized the structures of copper ion adsorption on rGO with zero, one, two, three and four oxygen functional groups, mainly considering the influence from oxygen functional groups at the first neighbor sites around the adsorption sites on rGO for simplification **(Supporting Information, Figure S5).** Figure 3B shows the representative optimized geometries of copper ion adsorption on both graphene and GO sheets that correspond to the unoxidized and oxidized regions of rGO respectively, as well as the relevant Mulliken charge distributions and highest occupied states of the molecular orbitals (HOMO) (see all optimized structures in Figure S5 in Supporting Information). We find that the binding energies of copper ion adsorption on rGO are all two orders of magnitude larger than the hydrogen binding energy in liquid water, indicating the strong adsorption of copper ions on rGO that is consistent with our detections in ICP-OES experiment (Supporting Information, Figure S1). Further, the HOMO represent all of the molecular orbitals with the cation-π interactions between the copper ion and rGO surfaces and clearly display the strong couplings between the orbitals of the copper ions and the delocalized π states of the aromatic-ring structures in rGO surfaces. [9a, 17] Mulliken charge distributions show that the copper ions adsorbed on rGO remain partial charges of around +0.5 in comparison with its original valence of +2, representing the significant charge transfers between the unoccupied valence orbitals of copper ion and the delocalized π states of the aromatic ring structure in the rGO. This demonstrates that the empty valence orbitals of the copper ion have a strong electron-withdrawing capability and accepted partial electrons from the delocalized π orbitals of the aromatic-ring structure of rGO during the adsorption process, consistent with our XPS experiments on the valence shift of copper ions from +2 to +1 after being adsorbed on rGO. All these results as well as the XPS experiments have demonstrated the strong cation–π interaction between copper ions and the aromatic rings of rGO such that copper ions can be adsorbed on rGO very stably.

Further, we performed MD simulation to analyze the interactions of the rGO–copper composite and the rGO with the bacterial cell membrane. [10a, 18] Herein, based on the cation-π interaction, we randomly selected sites on rGO and constrained copper ions onto rGO to form a rGO–copper composite (Supporting Information, Figure S6); meanwhile, bacterial cell membranes were simulated with a mixture of neutral palmitoyloleoylphosphatidylethanolamine (POPE) and negatively charged palmitoyloleoylphosphatidylglycerol (POPG) at a 3:1 ratio in explicit solvent environment,[10a] since negatively charged membranes usually occur with bacterial cells. [19]

Two representative MD trajectories of the rGO–copper composite and rGO interacting with bacterial cell membranes were shown in **Figure 4A** and Figure 4B, respectively. The rGO–copper composite is rapidly adsorbed and comes into contact with the membrane within 15 ns. By contrast, the rGO first moves around for ~115 ns; subsequently, after one of its corners becomes close to the membrane at ~115 ns, it takes an additional ~30 ns to come into complete contact with the membrane. The rGO–copper composite reaches the membrane much more rapidly than does rGO. Figure 4C further illustrates the movement of the centers of mass of the rGO–copper composite and rGO toward the membrane and presents their interaction energy profiles. As they approach the membranes, the rGO–copper composite exhibits a much more rapid increase in interaction energy with the cell membrane than does rGO, and its final energy is more than three times higher than that of rGO. This much higher energy essentially results from the electrostatic interaction between the copper ions of the rGO–copper composite and the negatively charged membrane. Therefore, rGO is functionalized to massively adsorb copper ions to form the rGO–copper composite through the cation–π interaction between copper ions and π-electrons in the aromatic rings of rGO. [9a, 17a] These copper ions on rGO are positively charged and strongly interact with negatively charged bacterial cells to achieve antibacterial activity, while rGO exhibits the functionality to actuate rapid delivery of copper ions and assembly onto bacterial cells to greatly enhance the antibacterial activity, which was consistent with our FE-SEM observations above (Figure 1C and 1D).

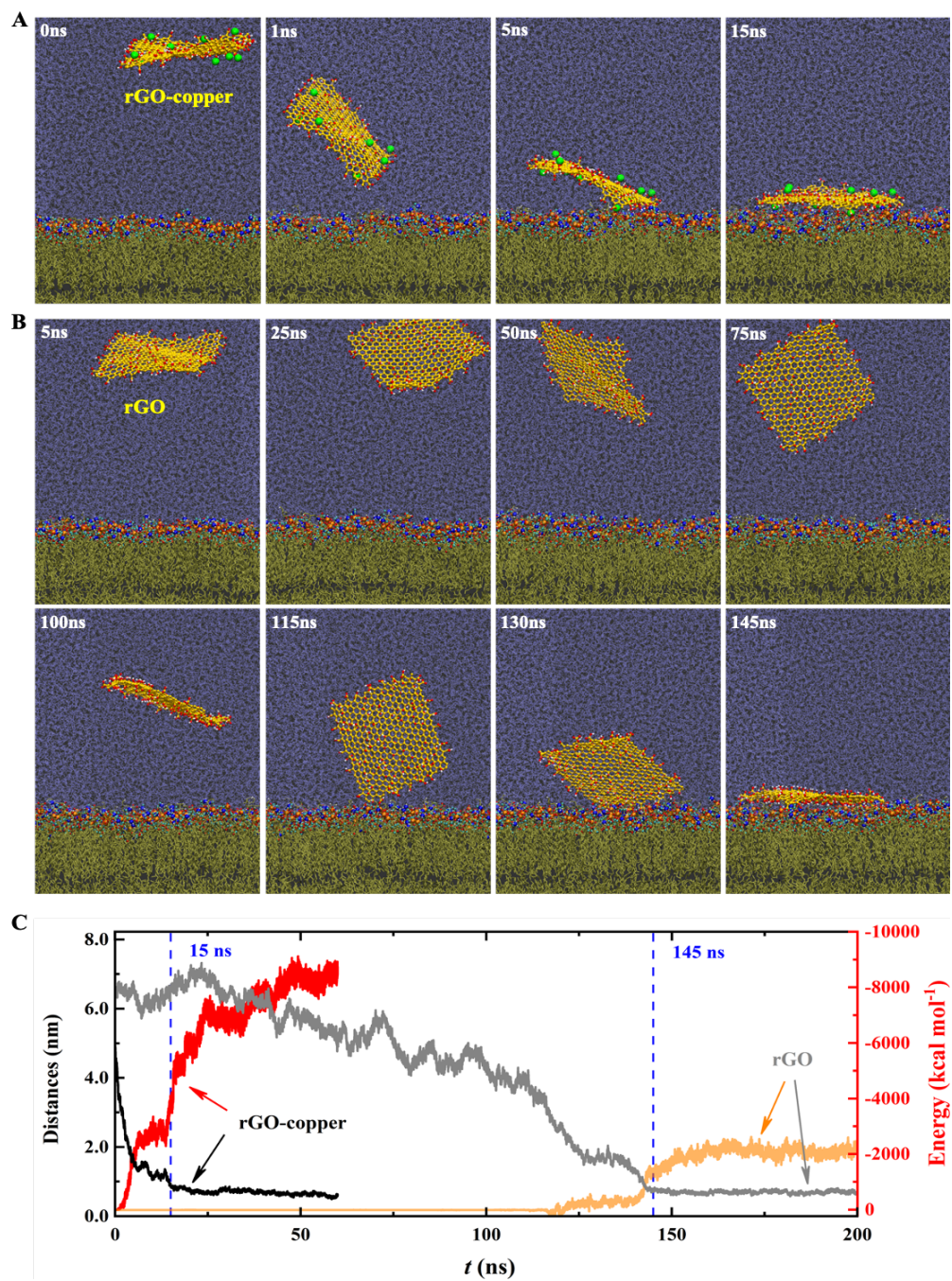

**Figure 4.** Molecular interaction of the functionalized rGO–copper composite with bacterial cell membranes. (A) and (B) Representative MD trajectories of the interactions between cell membranes (3:1 mixed POPE and POPG, negatively charged) and rGO–copper composite (*A*) and rGO (*B*). The snapshot times are displayed in the top left corners. Water is shown as ice blue lines, phospholipids as tan lines, and hydrophilic charged atoms as colored spheres (H: white, O: red, N: dark blue, C: cyan, and P: orange). The rGO is shown as bonded spheres with oxidized groups (C: yellow, O: red, and H: white). The copper ions of the rGO–copper composite are displayed as large green spheres. (C) Time evolution of the interaction energy between cell membranes and the rGO–copper composite and rGO (red and orange curves, respectively, right axis), and center-of-mass distances of the rGO–copper composite and rGO along the z-direction toward cell membranes (black and gray curves, respectively, left axis). The blue dashed lines indicate 15 and 145 ns, corresponding to the time at which the rGO–copper composite and rGO reached the membrane surface, respectively.

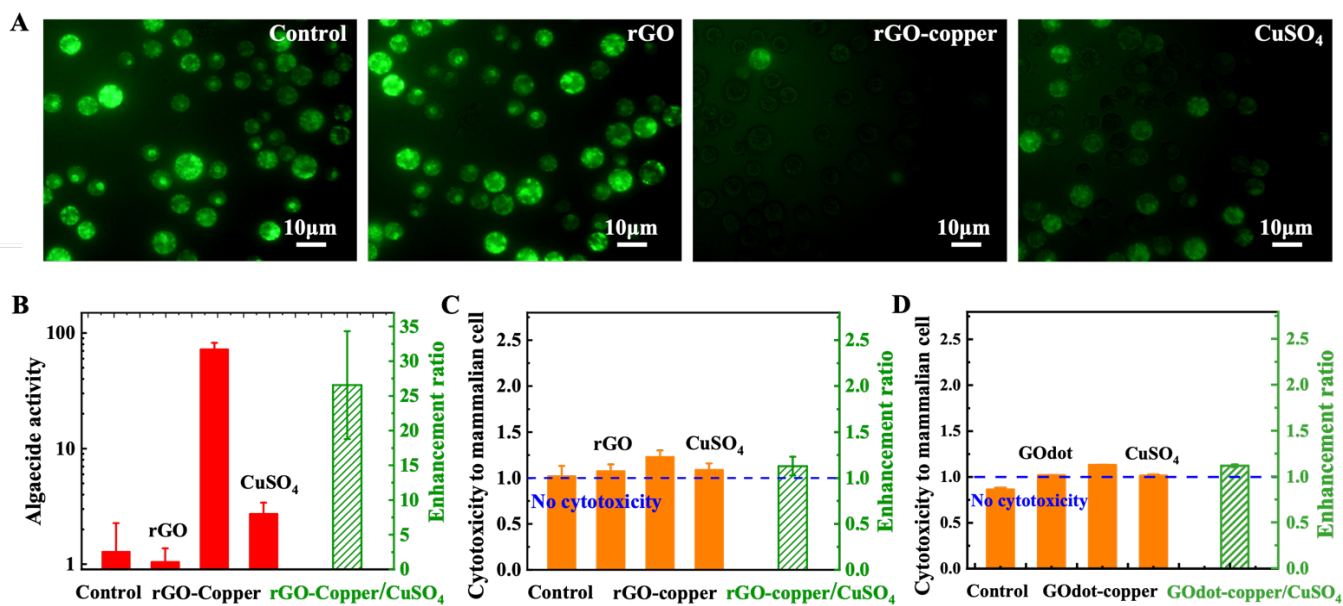

**Figure 5.** Algaecidal activity of the functionalized rGO–copper composite and its cytotoxicity against a mammalian cell. (A) Images of fluorescence-viable cells in experimental FDA assay of algaecidal activity of the rGO–copper composite against *C. reinhardtii*. (B) Algaecidal activity (left axis) of rGO–copper composite, and enhancement ratio relative to the surrounding copper ions (right axis). The error bars indicate standard deviations in 30 parallel experiments. (C) and (D) Cytotoxicity of rGO–copper (C) and GOdot–copper (D) composite against a mammalian cell (human embryonic kidney 293T), and enhancement ratios relative to the surrounding copper ions. The blue dashed line marks the level of no cytotoxicity. The error bars indicate standard deviations in three parallel experiments. The algae and mammalian cells were incubated with the same corresponding copper ion and/or rGO concentrations as in Figure 1A. The definitions of algaecidal activity and cytotoxicity and their relevant enhancement ratios are the same as those in Figure 1B.

**2.5 Algaecidal Activities and Mammalian Cytotoxicity of rGO–copper Composite.**

In addition to bactericidal activity, we verified the algaecidal activity of the rGO–copper composite. In fluorescein diacetate (FDA) assay, rGO-incubated algae (*Chlamydomonas reinhardtii*) cells are present as dense FDA-dyed viable cells similar to those observed in the control experiment (**Figure 5A**), indicating that rGO hardly present algaecidal activity. By contrast, after incubation with the rGO–copper composite, only one FDA-dyed viable cell is observed, indicating that the algaecidal activity of the composite is much stronger than that of rGO itself. The algaecidal activity of the rGO–copper composite is approximately 26 times that of the surrounding copper ions (Figure 5B).

For biomedical applications, examining the cytotoxicity of the rGO–copper composite against mammalian cells is critical. As shown in Figure 5C and 5D, all the rGO, rGO–copper, GOdot, and GOdot-copper composites have no cytotoxicity against a mammalian cell (human embryonic kidney 293T) **(Supporting Information, Table 1)**. The rGO–copper and GOdot–copper composites present almost the same cytotoxicity (only ~1.2 times stronger) as the surrounding copper ions, indicating no enhancement effect from the functionality of rGO and GOdot through copper ions. Previous cytotoxicity tests showed that only high copper ion concentrations (>$10^3$ μM) could damage mammalian cells. [20] In this study, despite the high local copper ion concentration on the rGO–copper composite due to the strong adsorption of copper ions by rGO, mammalian cells are usually neutrally charged [21] so that the copper ions of the rGO–copper composite are not preferentially delivered and assembled onto the mammalian cells; thus, the cytotoxicity is not enhanced. By contrast, algae cells usually have negatively charged membranes, similar to bacterial cells; [22] the algaecidal activity of the composite is explained by the same interaction mechanism as that for antibacterial activity.

## 3. Conclusions

In summary, we have demonstrated the remarkable selective antibacterial activity of the rGO functionalized by copper ions and no toxicity to mammalian cells. Through the functionality of rGO, the concentration of surrounding copper ions in solution is maintained at the extremely low level (less than approximately 0.5 µM, Supporting Information, Figure S1), which is at the same order of magnitude of water quality criteria. [14] More excitingly, the consequent rGO–copper composite exhibits antibacterial activity two orders of magnitude higher than that of its surrounding copper ions. Notably, this functionality of rGO through copper ions is also validated to similarly achieve algaecidal activity. In contrast with copper ions, we have also attempted to functionalize rGO by zinc and nickel ions but found no enhancement of antibacterial activity from the consequent rGO–zinc and GOdot–nickel composites **(Supporting Information, Figure S8)**.

In order to reveal the mechanism of this remarkable antibacterial activity, element copper mapping by FE-SEM has been employed. Cu contents on rGO–copper-incubated cells are approximately 20-fold larger than the increments on the $CuSO_4$-incubated cells, demonstrating that the rGO may prefer to massively deliver and assemble the adsorbed copper ions onto bacterial cells to enhance the antibacterial activity. Meanwhile, our DFT calculations indicate that this functionality of rGO originates from the strong cation-π interaction between copper ions and the aromatic rings of rGO **(Supporting Information, Figure 3B and Figure S5)**. Nearly all bacterial cells found in nature are negatively charged [19] **(Supporting Information, PS8)**, while the rGO are positively charged by copper ions through their cation–π interaction. Whether the oppositely electrical properties will enhance the interaction between rGO–copper composites and cell membranes has been explored by our MD simulations (Figure 4C and Supporting Information, Figure S7). The more rapidly contacting of rGO–copper composite to cell membranes as well as the much stronger interacting energy with the membrane in comparison with rGO only have confirmed our suspicions on the great enhancement on the interaction between positively charged rGO and negatively charged cell membranes. On the other hand, our XPS experiments and DFT calculations has further demonstrated that, in rGO-copper composite the rGO induces a valence shift of copper ions from +2 to +1, besides the strong adsorptions of copper ions (Figure 3 and Supporting Information, Figure S5). Notably, while the precise mechanism remains unclear, for more than 40 years it has been recognized $Cu^+$ ions are considerably more toxic to bacteria than $Cu^{2+}$ ions under test conditions where the enhancements of $Cu^{2+}$ cytotoxicity are mainly attributed to the reduction of copper that results in a shift in total copper from the $Cu^{2+}$ to the $Cu^+$ oxidation state. [23] Therefore, all these results reveal the inherent physical mechanism on the functionality of rGO through cation-π interaction with copper ions that not only actuates rapid delivery of copper ions and massive assembly onto bacterial cells but also results in the valence shift in the copper ions from $Cu^{2+}$ into $Cu^+$, which greatly enhances the antibacterial activity. In contrast, similar cytotoxicity test to HEK 293T cells has been accomplished by MTT assay, and both rGO only and the rGO functionalized through copper ions showed no cytotoxicity against mammalian cells.

It is worth noting that the GOdot, as smallest-lateral size GO, has also been demonstrated to have consistent functionality with the rGO to selectively achieve remarkable antibacterial activity yet exert no toxicity to mammalian cells. Our analysis supports that the lateral sizes have no essential influence on the rGO functionality. Meanwhile, the small lateral sizes of GOdot bring about the significant advantage on the high-probability interactions with bacterial cells to further boost up the antibacterial activity, since the number concentration of GOdot is drastically increased at the same mass concentrations as rGO. Nevertheless, despite the significant advantage, since it is rather hard to remove nanosized GOdot by the centrifugation or filtering, we may also have to reappraise and confront the environmental hazards of nanosized GOdot in practical use that is similar to the hazards from copper ions, and the choice of specific sized GO should be determined with specific application scenarios.

Herein, this functionality of rGO through cation–π interaction with copper ions was first revealed in our computations, and then was validated with our ICP-OES detections, FE-SEM observations, and XPS measurements. We attribute the selective cytotoxicity against both bacteria and algae cells to this functionality of rGO, and these copper ions on rGO are positively charged and selectively and strongly interact with both negatively charged cells rather than with neutrally charged mammalian cells. We note that pathological cells (e.g., cancer cells) likewise carry a net negative charge [24] and several cationic antimicrobial peptides were reported to display the advantage of their electrostatic interactions for cancer cell-selective toxicity. [25] The rGO functionality exhibited here takes the most advantage of the charge difference between the cells to achieve the remarkable selective antibacterial activity and meanwhile avoid environmental hazards, which offers guidance on the design for highly effective and selective anticancer activity. Considering the scalable production and the easy removal of rGO through centrifugation or filtering, the antibacterial and algaecidal activities from the functionality of rGO through copper ions as well as the inherent broad-spectrum-antibacterial physical mechanism represent a large step toward achieving a cheap, facile, highly efficient, broad-spectrum, and environmentally friendly antibacterial material and reagent for practical application.

## 4. Experimental Section

*Fabrication of reduced graphene oxide suspensions*: Graphene oxide (GO) suspension were prepared from natural graphite powder via a modified Hummers method [9a, 26]. Graphite powders were pre-oxidized by concentrated $H_2SO_4$, $K_2S_2O_8$, and $P_2O_5$ solution with continuous stirring for several hours. Then, we diluted the mixture with deionized water, centrifuged at 10,000 rpm, and washed with deionized water; after drying process, we obtained the pre-oxidized graphite. Further oxidization treatments were performed in concentrated $H_2SO_4$ and $KMnO_4$, diluted with deionized water, followed by the addition of 30% $H_2O_2$. In order to remove ion species, the product was centrifuged and washed sequentially with 1:10 HCl aqueous solution and deionized water. Eventually, by centrifugation at 4000 rpm, few-layer graphene oxide was separated to obtain GO suspension.

Electron-beam irradiation (EBI) reduction method is essentially radiation water production hydrated electrons as a medium involved in chemical reactions [27]. Hydrated electrons chemically react with the oxygen-containing groups on GO surfaces to reduce the oxygen content to achieve reduced GO (rGO) [28]. The GO suspension and isopropanol were mixed at a volume ratio of 3:2. The mixed solutions were put into a sealed bag filled with nitrogen gas and irradiated with a dose of 5 kGy under room temperature. The electron accelerator has a beam intensity of 5 mA and an energy of 1.8 MeV. Finally, the rGO solution was washed sequentially with deionized water by centrifugation.

*Purchase of graphene oxide quantum dots suspensions*: The GOdot suspensions (XF091, ~10 nm) were purchased from Nanjing XFNANO Materials Tech Co. Ltd.

*Atomic force microscopy (AFM) measurements*: AFM measurements were performed on Multimode V atomic force microscope, in tapping mode. The tips used have resonance of approximately 300 kHz. The samples for the AFM imaging were prepared by drop-casting a diluted suspension (0.03 mg/mL) onto a cleaned mica substrate and dried at room temperature.

*XPS measurements*: XPS was used to characterize the rGO–copper composite. The XPS measurements were performed with a ThermoFisher ESCALAB 250Xi spectrometer, using a monochromatic Al K Alpha (1486.6 eV) X-ray source (6 mA, 12 kV). We drop-casted the droplets of rGO suspension onto a smooth paper substrate to produce freestanding rGO membranes. These freestanding rGO membranes were dried at 70 ºC for 12 hours. Then, they were peeled off and immersed in $CuSO_4$ solution with a concentration of 1.05 M at room temperature for 2 hours. At last, the wet membranes with salt solution were taken out, remove the free solution by absorbent paper and dried again at 70ºC for 12 hours. The consequent dried membranes of rGO–copper composite were characterized by XPS. Meanwhile, pure rGO membrane, $CuSO_4$ and $Cu_2SO_4$ were also measured by XPS as the references for the membranes of rGO–copper composite. The anhydrous cupric sulfate ($CuSO_4$, 99%) powders were purchased from Shanghai Aladdin Bio-Chem Technology Co., Ltd. The anhydrous cuprous sulfate ($Cu_2SO_4$) powders were prepared with cuprous oxide ($Cu_2O$) and dimethyl sulphate ($C_2H_6SO_4$). Herein, 14.3g $Cu_2O$ powders and 10 ml $C_2H_6SO_4$ were added into a beaker, which were sealed with a PE film. The mixture was heated for 3h at 160ºC in oil bath; then, the products were dried in a vacuum drying oven.

*Measurements of surrounding copper ions concentration in composite solutions of rGO and copper ions*: The rGO suspension was directly added to $CuSO_4$ solution to achieve the rGO–copper composite solution. The copper ions being adsorbed by rGO was investigated at a temperature of 293 K. The rGO suspensions were added with a volume ratio of 1:1 into the $CuSO_4$ solutions, respectively. The mixed solutions were shaken at 180 rpm for 1 hour in 293 K constant temperature bath. These rGO–copper composite solutions were vacuum filtered through a microfiltration filter with a pore size of 0.22 μm (GREEN MALL, China) to remove rGO. The concentration of copper ions was determined with PS7800 inductively coupled plasma optical emission spectrometer (ICP-OES, detection limits 0.01~200 ppm (mg $L^{-1}$)).

*Cell strains*: *Escherichia coli* (*E. coli*) DH5α was used in this work as Gram-negative bacteria model, and *Staphylococcus aureus* (*S. aureus*) CMCC, as Gram-positive bacteria model. *Chlamydomonas reinhardtii* CC-125 was from Guoliang Xu's Lab (Shanghai Institute of Biochemistry and Cell Biology, CAS in China), used as algae model. Human Embryonic Kidney 293T (HEK293T) was used as mammalian cell model.

*Antibacterial activity enhancement test against E. coli and S. aureus by colony forming assays*: Antibacterial activities of rGO solution, $CuSO_4$ solution and rGO–copper composite solution were evaluated using *E. coli* and *S. aureus*. The bacteria cells were cultivated in Luria-Bertani (LB) medium (1% w/v tryptone, 0.5% w/v yeast extract, and 0.5% w/v NaCl) under shaking at 200 rpm at 37ºC for 16 hour (h) until it reached stationary growth phase. The bacteria cells were harvested by centrifugation at 4500 rpm for 5 min and washed two times with deionized water, and finally re-suspended in deionized water. Then, 500 μL of bacteria cells suspensions were introduced to 99.5 mL solutions, including deionized water (as a control), rGO solution (40 μg $mL^{-1}$), $CuSO_4$ solution (3 μM) or rGO–copper composite solution (rGO, 40 μg $mL^{-1}$, and $CuSO_4$, 100 μM), with a cell concentration of approximately $10^8$ $mL^{-1}$ (for *E. coli*) or $10^{10}$ $mL^{-1}$ (for *S. aureus*) based on their optical densities ($OD_{600nm}$). The 3 μM $CuSO_4$ solution is postulated to be a reference of the surrounding copper ions in the rGO–copper composite solution. These samples were incubated at room temperature under continuous shaking at 200 rpm for 2 h for antibacterial testing. To quantify the number of viable bacteria cells, each sample was diluted with standard serial dilution (by $10^4$ folds for *E. coli* and $10^6$ folds for *S. aureus*), and 500 μL of the suspension from each sample was uniformly spread on the LB agar plates and the number of viable bacterial colonies was counted after incubation at 37ºC for 24 h. Three parallel experiments with bacterial cells incubated independently for antibacterial testing were performed for checking reproducibility, and in each parallel experiment, three samples were adopted to evaluate the average number of viable bacterial colonies and its standard deviations. The antibacterial activity is calculated as the ratio of the colony number in the control experiment divided by the colony numbers in antibacterial experiments. The enhancement ratio of antibacterial activity is calculated as the ratio of the rGO–copper antibacterial activity divided by the antibacterial activity of its surrounding copper ions.

Antibacterial activities of GOdot solution, CuSO$_4$ solution and GOdot–copper composite solution were evaluated using *E. coli* and *S. aureus*. The above similar test procedures were carried out, with GOdot solution (40 µg mL$^{-1}$), CuSO$_4$ solution (3 µM) or GOdot–copper composite solution (GOdot, 40 µg mL$^{-1}$, and CuSO$_4$, 100 µM), and with a cell concentration of approximately 10$^8$ mL$^{-1}$ (for both *E. coli* and *S. aureus*). To quantify the number of viable bacteria cells, each sample was diluted with standard serial dilution by 10$^6$ folds, and 500 µL of the suspension from each sample was uniformly spread on the LB agar plates.

Antibacterial activities of ZnSO$_4$ and rGO–zinc composite as well as NiSO$_4$ and rGO–nickel composite were evaluated using *E. coli*. The above similar test procedures were carried out with the rGO solution (40 µg mL$^{-1}$), ZnSO$_4$ or NiSO$_4$ solution (3 µM), and with a cell concentration of approximately 10$^8$ mL$^{-1}$. Antibacterial activities of ZnSO$_4$ and GOdot–nickel and the antibacterial activities of NiSO$_4$ and GOdot–nickel were evaluated using *E. coli*. The above similar test procedures were carried out with the GOdot solution (40 µg mL$^{-1}$), ZnSO$_4$ or NiSO$_4$ solution (3 µM), and with a cell concentration of approximately 10$^8$ mL$^{-1}$. To quantify the number of viable bacteria cells, each sample was diluted with standard serial dilution by 10$^6$ folds, and 500 µL of the suspension from each sample was uniformly spread on the LB agar plates.

*Algaecide enhancement test against Chlamydomonas reinhardtii by fluorescein diacetate (FDA) assays*: Algae (*Chlamydomonas reinhardtii*) cells ware incubated in TAP medium in a biochemical incubator at 25 ºC under 12 h dark/lit illumination condition for 3 days until it reached exponential growth phase. The algae cells were harvested by centrifugation at 3000 rpm for 5 min and washed two times with fresh TAP medium. The cells were re-suspended in TAP medium itself as control, in TAP medium with rGO (40 µg mL$^{-1}$), CuSO$_4$ (3 µM), rGO–copper composite (rGO, 40 µg mL$^{-1}$, and CuSO$_4$, 100 µM), with a cell concentration of approximately 10$^6$ mL$^{-1}$ based on the optical density OD$_{750nm}$. The 3 µM CuSO$_4$ solution is postulated to be a reference of the surrounding copper ions in the rGO–copper composite solution. After 30 min incubation, the cells were washed twice with fresh TAP medium by centrifugations at 3000 rpm for 5 min and again suspended in TAP medium. Then, for cell staining, 1 mL of algae cell suspension was mixed with 1 µL of FDA stock solution in acetone (2% w/v) and incubated for 5 min at room temperature. Finally, the cells were washed twice and re-suspended in 1 mL of TAP medium. The viable cells were identified by the green fluorescent fluorescein which cellular esterases in the viable cells metabolize FDA and release, since dead cells do not cleave FDA and are distinguished by red fluorescence emitted from chlorophyll. Thus, 20 µL of cells suspension from each sample was uniformly spread on siliconized glass cover slide (FAstal BioTech). We used an inverted confocal microscope system equipped with a 60x immersion oil objective (Olympus PlanApo 60x; Cargille Type DF ImmersionOil) and CCD camera (sCOMS, 1270*1030 pixels, quantum yield=0.76, Dayueweijia (Beijing) Technology Co.,Ltd) equipped with Semrock 570/70 bandpass filter to take fluorescence pictures for counting the number of viable cells. The green fluorescence was emitted when excited at 488 nm laser with 30 µW intensity. 30 parallel experiments with algae cells incubated independently for algaecide testing were performed to evaluate the average number of viable cells and its standard deviations. The algaecide activity is calculated as the ratio of the viable cells number in the control experiment divided by the viable cells numbers in algaecide experiments. The enhancement ratio of algaecide activity is calculated as the ratio of the rGO–copper algaecide activity divided by the algaecide activity of its surrounding copper ions.

*Cytotoxicity test to HEK 293T cells by 3-(4,5-Dimethylthiazol-2-yl)-2,5-diphenyltetrazolium bromid (MTT) assay*: HEK293T cells were grown at 37ºC with 5% CO$_2$ in dimethyl sulfoxide (DMEM, Dulbecco's modified eagle medium, Sigma, Poole, UK) for 2 days to their logarithmic phase. The cells were harvested by centrifugation at 5000 rpm for 5 min and adjusted the cell concentration to 10$^4$~10$^5$ mL$^{-1}$ (counted with hemocytometers). Then 100 µL cells were added into every well of the 96-well plates and incubated for 2 h until all the cells adhered to the wall, and then the supernatant was removed and replaced by fresh DMED mediums itself as control, or by DMED mediums with rGO (40 μg mL$^{-1}$), $CuSO_4$ (3 μM) or rGO–copper composite (rGO, 40 μg mL$^{-1}$, and $CuSO_4$, 100 μM). After 4 h, the supernatant was removed and replaced by 100 μL of PBS, and then, 20 μL of MTT (5 mg mL$^{-1}$ in DMEM) was added into each well and the cells were incubated at 37 °C for another 4 h to form formazan. Finally, the supernatant with MTT was removed again, and 150 μL of DMSO was added to each well and the cells were incubated for 30 min to dissolve the formazan. The optical density (OD) at 490 nm were monitored by Synergy™ 2 Multi-Mode Microplate Reader (BioTek), and the cytotoxicity is calculated as the ratio of the $OD_{490nm}$ in the control experiment divided by the $OD_{490nm}$ in cytotoxicity experiments. The enhancement ratio of cytotoxicity is calculated as the ratio of the cytotoxicity of the rGO–copper composite divided by the cytotoxicity of its surrounding copper ions. Three parallel experiments with HEK 293T cells incubated independently for cytotoxicity testing were preformed to evaluate the average $OD_{490nm}$ and its standard deviations.

*Scanning Electron Microscopy (SEM) and Energy Dispersive Spectrometer (EDS) mapping images*: The *E.coli* cells (~10$^4$ mL$^{-1}$) were suspended specifically in deionized water, rGO solution (40 μg mL$^{-1}$), $CuSO_4$ solution (3 μM) or rGO–copper composite solution (rGO, 40 μg mL$^{-1}$, and $CuSO_4$, 100 μM) and then cultured at room temperature for 30 min. 100 μL of cells was dropped on the glass slide and dried by infrared heat lamp. The cells were sputter coated with platinum (20s, 20 mA) and then imaged with an SEM (LEO 1530vp) under 15 kV energy level radiation to analyze the morphology and EDS profile of cells. The element carbon and copper contents were quantified with the intensity of the peak at 0.27 keV and 0.93 keV respectively.

*Computer simulation methods*: In the DFT calculations, the structures of rGO with copper ion being adsorbed (rGO–copper) and the structure of rGO were optimized at the B3LYP/6-31+G(d,p) level of theory.[16] The geometry optimizations were performed via the Berny algorithm [29] with the convergence criteria of a maximum step size is 0.0018 au. and a root mean square (RMS) force is 0.0003 au. The visualized molecular orbitals were derived from the Kohn–Sham formalism and shown as an iso-surface value of 0.005 a.u. These calculations were carried out with the Gaussian 09 software package. [30] The binding energy of the copper ion adsorbed onto rGO is calculated by a formula of $\Delta E = E_{rGO–copper} − E_{copper} − E_{rGO}$, where $E_{rGO–copper}$, $E_{copper}$ and $E_{rGO}$ are the total energies of optimized rGO–copper composite, copper ion and rGO, respectively.

We performed MD simulations for both rGO and rGO–copper composite interacting with *E. coli* membranes. The simulation systems and a configuration of rGO-copper composite were shown in Figure S6 in Supporting Information. The rGO sheet (3.92 nm × 2.84 nm) were constructed based on the Shi-Tu structure model. [8a, 10a] The detailed parameters of rGO force field can be referred to previous publications. [10a, 18, 31] 10 sites on the rGO sheet were randomly selected to constrain 10 copper ions to model rGO–copper composite based on their cation–π interactions (Supporting Information, Figure S5). At each selected site, copper ion was constrained to its nearest carbon atom by a harmonic potential with the spring constant of 50 kcal mol$^{-1}$ Å$^{-2}$, approximately at the same order magnitude of the adsorption energies (Supporting Information, Figure S5). The negatively charged cell membrane that usually occurs in bacterial cells (for example, *E. coli* [19]) was modelled by a mixture of neutral POPE and negatively charged POPG at a 3:1 ratio. [32] In MD simulations, the GROMOS lipid force field (Berger) for POPE and POPG was adopted, which has been validated extensively. [33] Water molecules were represented by the simple point-charge SPC/E model. The periodic boundary conditions were applied in all directions. MD simulations were performed with Gromacs 4.5.7 software Package. The NPT ensemble is used with a time step of 2 fs, the temperature is maintained at 310 K by a Berendsen thermostat with a coupling coefficient of $\tau_T = 0.1$ ps, and the pressure is kept at 1 bar by the Berendsen barostat with a coupling coefficient of $\tau_P = 1$ ps. The particle-mesh Ewald method was adopted for the long-range electrostatic interactions with a cutoff of 1.2 nm, whereas a cutoff of 1.2 nm was also applied to the van der Waals interactions. The solvated lipid systems were initially equilibrated for 40ns. Then, the rGO sheet or rGO-copper composite was introduced into the system by replacing overlapping water molecules, and position restraints were applied to the rGO sheet and rGO-copper composite for 20 ns to allow systems re-equilibration. After the equilibration processes, the position restraints on the rGO sheet or rGO-copper composite were released and the consequent systems were adopted to continue to run MD simulations for analyzing both rGO sheet and rGO-copper composite interacting with the membranes.

## Supporting Information

Supporting Information is available from the Wiley Online Library or from the author.

## Acknowledgments

We thank Jianjun Jiang for preliminary testing calculations. The supports from NNSFC (Nos. 11675138, U1832150, 21773039), the National Science Fund for Outstanding Young Scholars (No. 11722548), the Special Program for Applied Research on Supercomputation of the NSFC-Guangdong Joint Fund (the second phase), Shanghai Municipal Science and Technology Major Project (No.2018SHZDZX01), SCI & TECH Project (No.20ZR1405800), and ZJLab.